# GearV: A Two-Gear Hypervisor for Mixed-Criticality IoT Systems

Kaiwen Long, Chong Xing, Yuebin Qi, Pei Zhang, Changsong Wu, Wenxiao Fang, Jing Tan,
Jie Chen, Shiming Zhang, Zuosheng Wang, Zuanmin Liu, Cao Liang, Jiaxiang Xu
Alibaba Group
{kaiwen.lkw, xingchong.xc, yuebin.qyb, yunzhu.zp, changsong.wcs,wenxiao.fwx, jing.tj,
cecil.cj, shiming.zsm, zuosheng.wzs, zuanmin.lzm, liangcao.lc, jiaxiang.xjx}@alibaba-inc.com

**Abstract**

With the rapid adoption of powerful computing platform in Internet of Things (IoT), such as industrial control and automotive scenarios, there is a strong demand to better utilize hardware resources while fulfilling the stringent real-time and reliability requirements regulated by certification. Embedded virtualization is an appealing technology to meet this demand through consolidating systems with different criticality levels on a single computing platform, such as best-effort systems and safety-critical systems. However, shared resource access on the consolidated platform could lead to resource contention which can easily jeopardize the timing predictability of real-time tasks for the safety-critical system. The static partitioning architecture pioneered by Jailhouse addresses the issue, but reveals limitation on resource utilization efficiency. The widely-used server grade hypervisors such as KVM and Xen have matured techniques for resource efficiency, but they were not designed with embedded constraints in mind. Meanwhile, the commercial embedded hypervisors are mature but closed-source and expensive. We argue that the main obstacle of adopting the open-source hypervisor technology in the mixed-criticality IoT systems is the monolithic hypervisor architecture, which cannot easily satisfy the requirement on stringent reliability.

This paper presents GearV, a two-gear lightweight hypervisor architecture to address the above challenges. By dividing hypervisor into partitioning mechanism and scheduling policy which is implemented in Gear1 and Gear2 respectively, GearV creates a consolidated platform to run best-effort system and safety-critical system simultaneously with managed engineering effort. The two-gear architecture also simplifies retrofitting the virtualization systems, because the partitioning mechanism in Gear1 is self-contained and fixed and thus upgrading Gear2 doesn't lead to the rigorous re-certification process. We have implemented GearV on a Renesas R-Car-H3 evaluation board. The result shows that the two-gear architecture reduces the code size by 1/3, the extra overhead introduced by Gear2 is negligible on CPU and memory virtualization, and 4.5% on I/O virtualization. GearV supports real-time VM with only 5% jitter overhead measured by Xenomai. We believe that GearV can serve as a reasonable hypervisor architecture for the mixed-criticality IoT systems.

## 1. Introduction

With the increasing computational power offered by modern embedded processors and its widely-adoption in IoT, such as industrial control and automotive scenarios, consolidating multiple operating systems and applications with different criticality level on the same computing platform demonstrates great value on reducing space, weight, power and cost. Thus, there is a strong demand to better utilize hardware resource while fulfilling the stringent real-time and reliability requirements regulated by certification. According to the levels of criticality [1][2], the computing system can be categorized to best-effort, mission-critical and safety-critical from low to high, and their reliability is $10^{-9}$, $10^{-6}$ and $10^{-4}$ failures per hour respectively. For instance, the reliability of cloud is $10^{-4}$ failures per hour while the reliability of automotive is $10^{-9}$ failures per hour. It requires that the safety-critical system follows the most rigorous design process and complies with the industry regulation, like IEC61508 and IS26262. The stringent reliability for safety-critical system and resource efficiency for best-effort system do not always match, thus favoring one may easily break the other. Embedded virtualization is an appealing technology to meet the requirements through enforcing strong spatial and temporal isolation while making the concurrent execution of systems with different criticality level on the same computing platform. However, the chosen virtualization solution must comply with the most stringent certification level required by the system. There are three architecture approaches to realize virtualization for mixed-criticality systems: hardware partition, software partition and dynamic scheduling.

The hardware partition solution is mainly provided by the SOC vendor. Examples of hardware virtualization include NXP i.MX8QP [9] and PowerVR Series6XT GPU [10]. NXP introduces a dedicated Corex-M4 called system controller unit (SCU) to manipulate the platform resource allocation in its i.MX8QP processor family. It can statically divide the ARM CPU core, GPU and memory into several partitions. Even that it is called hardware partition, there is still a software layer called system controller firmware (SCFW) running in the SCU to manipulate the resource partition policy [9].The hardware partition solution relies on the multiple hardware instances instead of time-sharing the same hardware unit, and depends on the firmware running on the propertied processor to manipulate the scheduling policy. Thus, it has advantage on strong isolation for security, but cannot achieve high resource efficiency due to the limitation of firmware.

The second architecture approach is the software partition which differs from the hardware partition on that it achieves spatial isolation through leveraging SOC's exposed standard hardware mechanism, such as X86 and ARM processor's VT extension. Intel ACRN [12] and Bao [13] rely the standard virtualization extension of X86 and ARM processor respectively. VOSYSmonitor utilizes ARM TEE extension to achieve the virtualization functionality [11]. These partitioning hypervisors can fulfill the stringent reliability requirement because they have very limited code footprint (< 20K SLOC) which could help ease certification, but the disadvantage is the resource underutilization obviously.

The third architecture approach is dynamic scheduling which is widely adopted in the server-grade hypervisor, such as Xen [14] and KVM [15] to achieve the resource efficiency with sophisticated scheduling policies. There are efforts to tailor these server-grade hypervisors to fulfill the stringent reliability requirement of mixed-criticality systems, but it is hard because of its architecture complexity and huge code size. QNX [16], Greenhill Integrity [17] and Open Synergy [18] provide commercial hypervisor technologies and claim that they support CPU and memory dynamic sharing. But there is no way to perform quantitative analysis against them because they are close-sourced.

We argue that system reliability and resource efficiency don't always match, and favoring one may break the other. System complexity hinders system reliability, but high resource efficiency relies on the complicated scheduling policy which contributes to the overall system complexity. We propose a two-gear hypervisor architecture called GearV, which help achieve resource efficiency while ensuring system reliability. The traditional hypervisor, no matter type 1 or type 2, puts all functionalities together. The basic idea of GearV to reduce the system complexity is to decouple the functionality of hypervisor into two components, including partitioning mechanism and scheduling policy. It introduces a simple hypervisor called Gear1 with small code size to implement the partitioning mechanism. The complicated scheduling policy is offloaded to Gear2 running in the primary VM of Gear1. In this way, the sophisticated scheduling algorithm can be tuned in Gear2 to achieve resource efficiency, while the system reliability is guaranteed by Gear1. We implement GearV by extending Google hafnium hypervisor [19]. The architecture and implementation of GearV to restructure the open-source hypervisor can also be ported to other open-source ones. In summary, the paper makes the following contributions.

- It analyzes the requirements of mixed-criticality systems to identify the functional decomposition of embedded hypervisor.

- It proposes the first two-gear hypervisor architecture and implements a reference called GearV.

- It conducts the comprehensive evaluations of GearV on embedded SOCs through micro-benchmarks and application benchmarks to compare its virtualization cost against the monolithic hypervisor architecture.

- It demonstrates that GearV not only guarantees the resource efficiency with negligible virtualization overhead, but also achieves the desired reliability through controlling system complexity.

The rest of the papers is organized as follows. The background and the problems to address are described in section 2. The design and implementation of GearV are presented in section 3. The evaluation results are discussed in section 4. The conclusion and future work are presented in section 5.

## 2. Motivation

### 2.1. The Problem

It is no doubt that the hardware partition approach is the most stringent one to achieve system reliability. The static partition approach is useful for the scenarios that both the physical resource and workload can be pre-defined. But more and more applications with different criticality level are consolidated on a single embedded SOC, thus the resource efficiency becomes one of the core competences for the cost-sensitive IoT systems. Time-sharing of the physical resource increases resource efficiency but leads to the side-band attacks, like Meltdown and Spectre [20]. Therefore, fulfilling the stringent reliability and resource efficiency requirements within the same hypervisor leads to a need of exploring a new architecture.

### 2.2. Existing Embedded Hypervisors

Previous studies generally tackle the embedded virtualization in several directions, such as tailoring server-grade hypervisor, re-inventing hypervisor with small code footprint, and extending RTOS with virtualization support.

First, with the great success of Xen and KVM in cloud world, there are efforts to port existing server-grade hypervisors to embedded world, such as RT-Xen [21] and embedded KVM [22]. This approach is appealing when workload consideration and software compatibility are preferred over stringent reliability. However, it cannot easily satisfy the real-time constraints from IoT scenarios. Meanwhile, the restructuring process to pass reliability certification like IEC61508 and ISO26262 involves with heavy source code modification and tailoring effort.

Second, there are some open-source type-1 hypervisors written from scratch, such as ACRN [23], Xvisor [24], Jailhouse [28], and so on. ACRN only supports x86 architecture and doesn't work for the ARM architecture. GPL-licensed Xvisor and Jailhouse are not friendly to intellectual property protection for the commercial use case.

Third, the proven-in-use RTOS is extended with virtualization capability. There are many commercial examples of this kind, such as OKL4 Microvisor [25], QNX hypervisor [16], Green Hills INTEGRITY Multivisor [17], PikeOS [26]. Normally these RTOSes are based on microkernel architecture, and the virtualization capabilities are implemented as RTOS service routines or modules. Thanks to their small lines of source code, it is more feasible to get industrial certification like ISO26262 than Linux. But they are expensive and close-sourced.

### 2.3. Google Hafnium Hypervisor

Hafnium [19] is Google's open-source type-1 hypervisor following BSD-3 license, which targets at isolation between trusted domains and untrusted domains. It only provides CPU and memory partition, and doesn't contain the device model for I/O virtualization. The architecture of Hafnium is similar to KVM's split-mode, and it relies on the primary Linux VM to schedule the secondary VMs. It only has around 20K lines of code, and can be restructured to get reliability certification, such as IEC61508 and ISO26262. We choose Hafnium as the code base of GearV, and restructure it with the guidance of two-gear hypervisor architecture.

## 3. Design and Implementation

We propose a two-gear hypervisor architecture called GearV which can fulfill the system reliability and resource efficiency of the mixed-criticality IoT systems. The functionality of an embedded hypervisor can be divided into two parts: partitioning mechanism and scheduling policy. The partitioning mechanism is usually simple and encapsulated into the Gear1 hypervisor as the root of system reliability. The scheduling policy is related to workloads and usually complicated. For the interests of simplifying Gear1, the scheduling policy is offloaded to Gear2 running in the primary VM of Gear1.

In this section, we present the detail design and implementation of GearV. First, we highlight the design goals of two-gear hypervisor architecture. Second, we introduce the virtualization extension of ARMv8-A architecture, followed by the spatial and temporal isolation. Third we describe the techniques to achieve device virtualization. Finally, we conclude this section with real-time and security support.

### 3.1. Goals

A modern embedded hypervisor should be designed with the following goals:

- **Managed Complexity**: The root cause that hinders the stringent reliability is system complexity. One important indicator of system complexity is the line of code (LoC). According to the industry best practice, the reasonable code size with managed complexity is around 10K LoC.

- **Minimized Overhead**: Since the IoT system is cost sensitive, it is important for the application to make full use of the available computing and memory resource. Therefore, it requires that the hardware virtualization extension to reduce virtualization cost should be leveraged in embedded hypervisors.

- **Extensibility for Rich Features:** The modern hypervisor should be extensible to support the available guest OS in the market and further the innovative applications. For instance, one challenge in IoT systems is about lively adjusting the memory threshold or physical cpu cores among different VMs. Image that two guest OSes are running on the same vehicle computing platform, one for navigation and another one for video playback, the available cpu and ram should empower the navigation OS during driving and then shift to the guest OS of video playback during parking.

- **Real-time:** The hypervisor should support real-time VM (RTVM), in order to fulfill the deterministic characteristic of IoT systems.

### 3.2. Two-gear Hypervisor Architecture

To achieve the managed complexity and hence being easy towards industrial certifications, GearV relies on the careful decomposition of the core hypervisor functionalities at architecture level. The hypervisor can be broadly classified as virtualization techniques on CPU, memory and device. Each has specific behavior and different hardware supports. Regardless of the classification, the functional role of a virtualization technique remains the same, namely, arbitration of platform resources, and seamless operation of individual guest OS with minimal porting effort and runtime sacrifices. In this way, the hypervisor functionality can be divided into partitioning mechanism and scheduling policy.

- **Partitioning Mechanism:** It refers to the ways in which CPU cores, memory and devices are partitioned at initialization stage and shared among VMs at runtime.

- **Scheduling Policy**: It refers to the runtime policies like vcpu scheduling, on-demand memory allocation and I/O virtualization implemented in a dedicated VM. But for some performance critical devices like virtual generic interrupt controller (vGIC), GearV implements them in Gear1 hypervisor for fast responsiveness.

The partitioning mechanism and scheduling policy are implemented in Gear1 and Gear2 respectively, as shown in Figure 1. Gear1 is designed to make use of the hardware virtualization available in ARMv8-A's EL2 mode to provide four basic mechanisms. First, Gear1 performs one time initiation on memory partitioning, low level interrupt routine setup and per physical CPU execution context, according to the platform configuration and VM configuration. Gear1 enforces protection and isolation between different VMs through manipulating the hardware protection features like stage-2 MMU. As the lowest layer to interact with the underneath hardware directly, Gear1 is highly critical and its code base is kept to an absolute minimal. Second, Gear1 implements a set of clearly-defined portable hypercall APIs like inter-vm-communication and virtual interrupt injection. Third, Gear1 performs the CPU execution context switch from the primary VM to the guest VM and vice-versa. An execution context, also referred as world, contains the general registers, system registers and also the address mapping table. Switching from one world to another one is called world switch. Fourth, Gear1 traps necessary exceptions for device emulation, either direct handling for performance (e.g., GIC emulation), or routing to GDM/Device Emulation VM for complex emulation (e.g. virtio block).

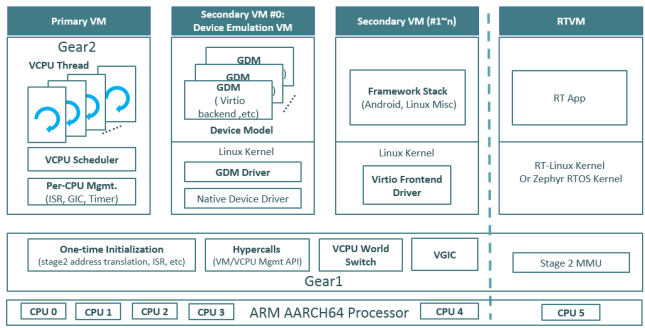

Figure 1 GearV Architecture Overview

Gear2 runs in kernel mode inside the primary VM of Gear1 hypervisor. It focuses on the vcpu scheduling of the secondary VMs and manipulates all the available CPU of the platform. Therefore, Gear2 can be divided into two layers. The lowest layer acts as a mini-OS which could handle physical interrupt, manipulate the multiple processors and abstract the thread scheduling framework. Above the mini-OS layer, an idle thread is created for each physical CPU and the vcpu thread is created for each vcpu. After all physical CPUs are powered up, each vcpu thread is kicked off through invoking the world switch hypercall to switch to the secondary VMs. From Gear2's point of view, world switch hypercall will block and only return after the secondary VM quitting the vcpu execution, with pre-defined reason. For example, while the secondary VM executes a MMIO operation, it traps into Gear1 which performs the world switch through saving the context of secondary VM and restoring the context of primary VM to Gear2. And then, Gear2 emulates the MMIO operation through injecting the virtual interrupt to the device emulation VM (DVM). After that MMIO emulation is done in DVM, it sends acknowledge back to Gear2 which then moves forward the instruction IP of the secondary VM. In this way, the sematic of MMIO operation of the secondary VM is emulated.

Due to the two-gear hypervisor architecture, it involves multiple mode transitions to/from Gear2 and leads to more traps than the traditional monolithic hypervisor. However, as shown in Section 5, these extra traps from/to Gear2 doesn't introduce significant performance cost on ARMv8-A.

### 3.3. Layered Reliability Supervision

The benefit of hypervisor is that functional disruptions in one VM cannot affect other VMs and the hypervisor is able to supervise whether VM works as expected. Using watchdog is a straightforward way to implement the reliability supervision mechanism. Thanks to the two-gear architecture, having separate Gear2 hypervisor to encapsulate the complex scheduling policy simplifies the challenge to achieve stringent reliability and can add one more layer of reliability supervision, as shown in Figure 2. Unusually, an external MCU monitors the rich SOC, as shown as L4RS. At the SOC side, besides the traditional two layered reliability architecture of L2RS and L1RS, GearV adds another layer called L3RS.

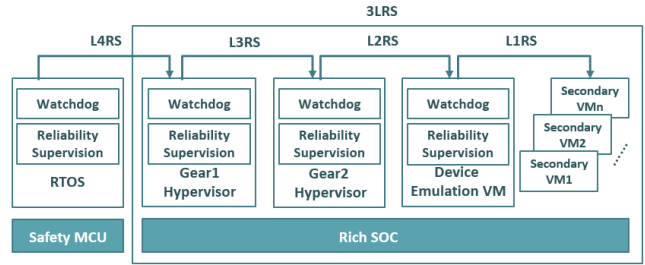

Figure 2 Layered Reliability Supervision Architecture

### 3.4. ARMv8-A Virtualization Extensions

To reduce the virtualization cost, the ARM architecture introduced hardware support as an optional extension in the latest ARMv7 and ARMv8. For example, the ARMv8-A architecture provides the following five categories of virtualization hardware support:

- **CPU Execution Mode**: A dedicated Exception Level (EL2) with dedicated registers is introduced for hypervisor.
- **Sensitive Operation Capturing**: It adds support to trap operations which access and manipulate sensitive state or resource in EL1 and EL2 mode.
- **Interrupt Routing**: It supports the mechanism that asynchronous interrupt and synchronous exception can be configured to trap into hypervisor in EL2 mode, and the virtual interrupt can be forwarded to a specific VCPU.
- **MMU Virtualization**: Two-stage memory address translation is supported and the second stage is for the hypervisor to isolate guest OSes.
- **Explicit instruction for hypervisor call (HVC).**

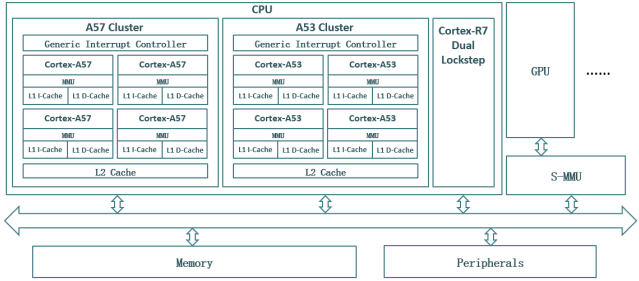

Figure 3 R-Car 3H SOC Diagram

An example of the ARMv8-A architecture is the R-Car 3H SOC illustrated in Figure 3 [30], on which GearV is implemented and evaluated. It employs the big. LITTLE architecture with Cortex-A53 cluster for low power and Cortex-A57 cluster for high performance. GearV is chosen to run with the AArch64 execution state to make full use of these hardware support features.

### 3.5. Spatial Isolation

Spatial isolation is the fundamental motivation of embedded virtualization. Running applications in a VM prevents bugs or malicious parts of the applications from interfering with other applications or data on the same embedded consolidated platform. Regarding to spatial isolation, all the shared hardware resources face two challenges on the partitioning policy and the corresponding interference prevention mechanism, which determine the resource efficiency and the degree of reliability. GearV implements spatial isolation mainly through partitioning memory, CPU core, cache and memory bus, and I/O devices.

#### 3.5.1. Memory Partitioning

The goal of memory partition is to provide memory separation among VMs. To achieve the goal, there are two tasks: physical memory region allocation, and address mapping from virtual space to physical space, as shown in Figure 4.

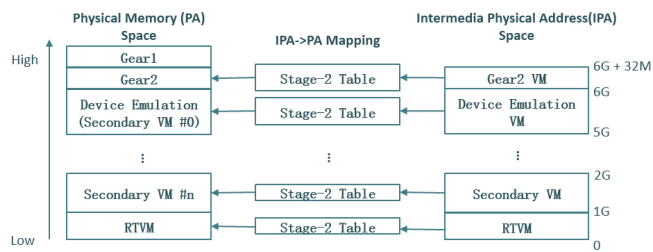

Figure 4 Memory Partition Layout

***Physical memory region allocation***: At bootup time, Gear1 retrieves the physical memory layout information from the flattened device tree (FDT) table passed from bootloader and divide them at page granularity. Gear1 reserves a subset of physical memory pages for its private use, and allocates mutually exclusive ones to guest VMs according to manifest.

***Address mapping from virtual space to physical space***: The memory access violation is detected through virtual to physical address mapping. Gear1 provides memory virtualization by leveraging stage-2 translation for all memory access when running in a VM. Stage-2 translation can only be configured in EL2 mode and is completely transparent to the VM. Gear1 manages the stage-2 translation page tables to allow access to physical memory regions only allocated for the VM, and other accesses will cause stage-2 page faults into Gear1 hypervisor. During the world switch from one VM to another one, Gear1 enables stage-2 address translation for the target VM.

For the interests of simplicity, Gear1 uses the pre-built page tables to hold the identity address mapping, such as stage-2 page tables for VMs and stage-1 page tables for itself. Thanks to the infinite address space for 64bit architecture, the IPA (intermediate physical address) is configured equal to the PA (physical address) in the stage2 page table, if it is mapped. There is a special case that the IPA of MMIO region is left as hole and unmapped in the tage-2 page table intentionally, in order to trigger trap for device emulation. Configured IPA range is passed to every VM through patching FDT table. By leveraging the virtualization hardware support on MMU, physical memory is partitioned at the initialization stage. The extra value of two-gear architecture on the memory partitioning is that more advanced optimizations could be conducted in Gear2 for memory utilization efficiency without breaking Gear1. For instance, Gear1 exposes the basic mechanism to manipulate the <ipa, pa> mapping, and Gear2 can dynamically adjust the available physical memory regions according to the runtime behavior of VMs, e.g., balloon driver.

#### 3.5.2. CPU Core Partitioning

Each physical CPU (PCPU) is self-contained since it owns private general registers, system registers, MMU, L1 cache, local interrupt controller and even local timer. The goal of CPU core partitioning is to bind the virtual CPU (VCPU) to PCPU and to guarantee that one VCPU context is not leaked to another one if they share the same PCPU. The ARMv8-A architecture introduces EL2 mode to run hypervisor code. Combined with the hardware support, Gear1 can simply save/restore the register state into/from its internal per-vcpu structure during world switch if the register doesn't affect VM isolation. The access or manipulation to all other sensitive states can be configured to trap into Gear1 for emulation. For example, the VM traps into Gear1 if it executes the WFI (Wait for Interrupt) instruction which causes the PCPU to power down and to wake up upon interrupt. Besides the consistent register state presentation, there are another two tasks regarding to CPU core partitioning: MP Bootup and VCPU binding.

*MP Bootup*. At bootup time, Gear1 parses the PCPU and cluster layout information from the FDT table. For the interests of simplicity, the multi-core brings up routine is offloaded into Gear2 and Gear1 just implements the mechanism of powering up a specific CPU core. Once retain control from bootloader on the primary PCPU, Gear1 performs one time initialization, such as setting up stage 2 translation table and per-vm initialization in memory, and then transfers control to Gear2. Then Gear2 performs its own one-time initialization, such as setting up low level exception table and vm/vcpu management structures, and finally calls the PSCI_CPU_ON hypercall to power up the secondary PCPUs. The PSCI_CPU_ON hypercall is captured by Gear1, which further issues call to the SOC vendor's firmware in EL3 mode. Each PCPU is registered with a pre-defined thread scheduler and enters into an idle thread loop after the PCPU powers up.

*VCPU Binding*. Gear2 is designed to support various VCPU scheduling algorithms. To simplify the system design, the local scheduling policy is implemented as the first step. A general pluggable thread scheduling framework is implemented in Gear2 and each PCPU is registered with a dedicated scheduler. The default one is a local round-robin thread scheduler, which maintains run-queue list of VCPU and each VCPU thread is executed in round-robin manner. When a VCPU thread is created, Gear2 puts it into its preferred PCPU's run-queue according to the VM's PCPU affinity configuration, and then kicks off the VCPU thread execution through calling the world switch hypercall. When some VM exit conditions are triggered in VM execution mode, Gear1 hypervisor gains control and forwards the control back to Gear2, which decides the following actions, such as how to schedule the next VCPU thread execution. In this way, Gear2 can simply bind VCPU to PCPU and other sophisticated VCPU scheduling policy can be plugged in according to the workload scenarios.

### 3.5.3. Cache Coloring

Ensuring predictable real-time performance is one of the key requirements for the safety-critical systems. The large last-level caches (LLC) can easily jeopardize the timing predictability of real-time tasks due to cache interference which is caused by contention on the LLC. The cache interference has been studied extensively in both real-time systems and virtualized ones [34][35].The technique to achieve spatial isolation on LLC is called cache coloring. It exploits the behavior of set-associative caches, which use part of the PA, denoted as set index, to identify the cache line to be used. The key rationale is that, if a VM accesses only PAs whose set indexes match a given pattern, then it will also access a restricted set of cache lines. In fact, cache coloring aims at reserving a subset of the bits composing the set index to identify a LLC partition, and is used to create the RTVM.

### 3.5.4. Device Partitioning

One important difference that embedded virtualization differs from the server virtualization is the I/O device. The typical usage case in embedded world is to run particular functionality with particular I/O device, thus static partition and pass-through are more preferred than resource migration and over-commit, such as VM migration and memory overcommit. There are three ways to partition I/O devices: pass-through, API forwarding and virtio[36].Pass-through is preferred for RTVM, API forwarding works for the case that there is standard API set to encapsulate the device, like OpenGLES and OpenCL for GPU, and Android HAL for camera sensor. Virtio is already the de-facto standard for the legacy device virtualization.

### 3.6. Temporal Isolation

One of key advantages of virtualization is the possibility to seamlessly pack multiple under-utilized systems into a single SOC, thus achieving a better utilization of the available hardware resources. But it adds a level of unpredictability in the performance that may be exhibited by each individual VM. For example, a VM with a temporary compute-intensive task might disturb the other running VMs, causing an undesirable temporary drop in their performance. The goal of temporal isolation is to ensure the correct distribution of the CPU time among VMs with different criticality level. For RTVM, GearV adopts the straightforward approach, which assigns separate dedicated CPU cores to RTVM. For the interests of simplicity, the device VM and secondary VMs are equally treated, and their VCPUs are scheduled by Gear2 using the round-robin scheduling policy at the beginning. Other scheduling strategies, such as BVT, CREDIT and EDF, will be explored in future.

### 3.7. Device Virtualization

*Background*: Different from that the server side I/O devices are usually standard, the I/O devices of embedded systems are highly diversified and tightly bound to a particular workload, we examine the device's intrinsic behavior and its programming model in the user's perspective, and prefer pass-through plus API forwarding for the emerging embedded device emulation like camera and GPU in GearV, for the interests of portability and performance. Of course, virtio and full emulation are also used to virtualize the legacy devices like ARM GIC, timer, disk and network controller. The major I/O devices supported on GearV and their virtualization techniques are summarized in Table 1.

Table 1 Device Virtualization Technique

| Devices | Technique |
| --- | --- |
| Audio/Codec/Camera/GPU | Pass-through, API forwarding |
| Network/Disk/Console/Touch | Pass-through, Virtio |
| GIC/Timer | Pass-through, Full emulation |

*API Forwarding*: There are already standardized API set for emerging devices, such as Android HAL for audio and camera, OpenMX IL for codec, OpenGL and OpenCL for GPU, but their implementation and even the document are proprietary and closed-source. Thus, rather than modeling a complete audio and or video processors, GearV chooses the pass-through and API forwarding approach to achieve device multiplexing. It assigns the devices to DVM which interacts with the underneath devices directly except for the physical interrupt delivery. Other access from secondary VMs through standardized API set is intercepted by the API wrapper library and then forwarded to GDM (GearV Device Model) running in DVM. In this way, secondary VM's access to the device is entirely mediated through the vendor provided APIs and drivers on DVM.

*Virtio*: The existing virtio frontend drivers of disk, serial console and touch panel are directly reused and the backend drivers are re-implemented as user space routines in GDM. The MMIO to bridge frontend driver and backend driver is trapped into Gear1. Furthermore, the I/O intensive module could be rewritten as kernel module to remove the running level switch.

*Full Virtualization:* Trap-and-emulate is used to virtualize ARM Generic GIC and timer. As an interrupt controller, ARM GIC exposes two kinds of programming interfaces: global distributor for global operation, and per-CPU interface for local CPU operation. From GICv2, ARM adds hardware support on GIC's per-CPU interface called vGIC, which is frequently accessed in the interrupt service routine. Gear1 just maps GIC's per-CPU virtual interface to a VM, allowing software in that VM to manipulate GIC directly.VM access to GIC's global distributor interface is still trapped and emulated, but such kind of operations often occur in initialization and inter-processor interrupt (IPI). In this way, the virtualization overhead on ARM GIC is nearly eliminated.

*Timer:* On ARMv8-A, each timer is per-cpu and consists of a set of comparators that compare against a common system count, and generates an interrupt when its value is equal to or less than the system count. Because multiple VCPUs can share the same PCPU and disabling timer interrupt by the online VCPU could lead to that the offline VCPUs loses its expected timer interrupt, even that they become online again. In GearV, EL1 virtual timer is exposed to guest VM directly and its related registers are mapped into per-vcpu state saved/restored during vcpu switch. EL2 physical timer is enabled in Gear1 to emulate the expected timer interrupt of the offline VCPUs. For example, the comparator of EL2 physical timer is set to the nearest value of offline VCPU's. Upon EL2 timer interrupt fires, Gear1 injects into vGIC with the pending virtual interrupt which is fired upon that the target VCPU retains control again.

*Interrupt Delivery*: The final topic regarding to device virtualization is the interrupt delivery mechanism. As shown in Figure 5, expect for the RTVM, the physical interrupt is configured to cause trap from VM to Gear1 and only Gear2 is allowed to handle physical interrupt directly. Through configuring system register (HCR), the interrupt is delivered to Gear2 directly, and trapped into Gear1 if it hits the other secondary VMs. Upon physical interrupt occurs in DVM and other secondary VMs, it conducts world switch to Gear2, which performs VCPU thread scheduling and then injects the virtual interrupt to the corresponding vcpu in DVM. In DVM, the low interrupt is transformed to I/O event which is handled by the user space GDM. After that I/O event is serviced in GDM, it might inject virtual interrupt to corresponding secondary VM according to virtio flow. The emulation process of MMIO operation is similar to the physical interrupt hitting a secondary VM. The difference is that the MMIO has both blocking and non-blocking semantics. For the blocking MMIO, the initiated vcpu should be blocked in the vcpu thread in Gear2 until the completion of the MMIO operation.

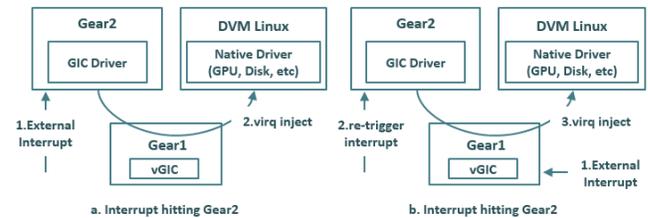

Figure 5 External Interrupt Delivery Flow

### 3.8. Real-time Support

Modern SOC is usually built with the multi-core processors which are designed to maximize throughput at the expense of latency and determinism, with unpredictable timing behavior such as cache and speculative execution. The major sources of unpredictability arise from inter-core contention over shared resources, such as LLC, memory controller, bus, interrupt and I/O devices. To the best of our knowledge, there are two approaches to tackle the unpredictability issue in the hypervisor: real-time scheduling and static partitioning. Even that real-time scheduling like RT-Xen [33] has advantage on resource efficiency, it can only work for the soft real-time system because of the interference on the micro-architectural resources like cache and memory bus. The static partitioning approach is widely used for hard real time in practice, because the architectural level partitioning such as CPU cores, RAM and I/O devices, and micro-architectural level partitioning such as cache, interconnects and memory controller, eliminate the interference from different guest VMs and further the unpredictability factor contributed by the hypervisor itself. How to ensure the predictability of the real-time tasks running on the SOC with unpredictable timing behavior is a challenge and out of the scope of this paper. We evaluate the real-time performance of a hypervisor through evaluating the scheduling latency jitter. For the interest of simplicity, the static partitioning is adopted in GearV.

To remove the unpredictability due to resource contention, Gear1 creates a special isolated guest VM called RTVM with dedicated CPU cores, memory and hardware devices through configuring the stage 2 memory translation table at the boot stage. The pass-through devices include the per-cpu timer, per-cpu GIC interface, disk and so on. At runtime, the RTVM accesses the allocated hardware resources directly without involving Gear2 and DVM. The only required VM exit from RTVM is the operation on GIC global distributor (GICD) which is used for setting up interrupt route mapping and IPI. The trap-and-emulate approach for GICD guarantees that the RTVM has no chance to attack the underneath hypervisor or other VMs, such as continuously broadcasting IPI to PCPUs belonged to other VMs. In this way, there is no resource contention on CPU and RAM with other VMs and RTVM is run inside a sandbox except the cache coloring and memory bus bandwidth reservation which is in development during this paper writing.

Static resource portioning for RTVM has obvious disadvantage on that the number of RTVM is limited by the availability of hardware resource, compared to the real-time scheduling, but the rationale is that the resource requirement for the real-time tasks is predefined and limited, and the engineering effort is managed at the acceptable level especially for the hard real-time guest system. We believe that static partitioning is the reasonable approach for embedded virtualization.

### 3.9. Security

GearV is built upon Hafnium and therefore inherits its security model by default. The design goal of Hafnium is to achieve confidentiality and memory integrity of each VM through memory isolation, following the Arm Platform Security Architecture Firmware Framework for Arm v8-A (PSA FF-A) specification [40]. It adds ownership state to each physical page, thus the physical memory can be shared between VMs explicitly through leveraging stage-2 address translation table. The mechanism to denote and share memory is reused in Gear1.

## 4. Evaluation

In this section we conduct evaluations of GearV according its design goals on managed complexity, virtualization overhead, extensibility and real-time support. After describing the hardware and software configuration, this section mainly aims to answer the following four questions:

1) *How does the system complexity of GearV compare to that of existing hypervisors like ACRN and Xen?*
2) *How does the detail virtualization overhead imposed by the two-gear architecture compare to that of monolithic architecture pioneered by Xen and KVM?*
3) *How about the overall performance impact of applications running on the GearV platform?*
4) *How does GearV behave regarding to the real-time characteristics?*

### 4.1. Configuration

Table 2 lists the hardware configuration for GearV evaluation. For hardware platform, an R-Car H3 Starter Kit Board with 4 ARM Cortex-A57 cores and 4 ARM Cortex-A53 cores is used. For software platform, Linux 4.14 is used for both device emulation VMA and the other secondary VMs. Two kinds of real-time enhancements including PREEMPT_RT and Xenomai 3.1 are used to evaluate the real-time performance of GearV. Table 3 lists the resource configuration for each VM, mainly including CPU core number and RAM budget.

Table 2 HW Configuration for GearV Evaluation

| Board | R-Car H3 Starter Kit |
|---|---|
| CPU Core | Cortex-A57 Quad, Cortex-A53 Quad |
| Memory | 8GB |
| Storage | 32GB |
| GPU | PowerVR GX6650 |

Table 3 Resource Configuration for Single VM

| VM | Dev-Emul VM | Secondary VM | RTVM |
|---|---|---|---|
| VCPU # | *3* | *2* | *1* |
| Memory | 3GB | *512MB* | *512MB* |
| Storage | pass-through | virt-io | pass-through |

To conduct comprehensive evaluation of GearV, both micro-benchmarks and application benchmarks are used. KVM is used as the reference hypervisor because its split-mode architecture is similar to the two-gear architecture of GearV and it can run successfully on the R-Car H3 board. The host VM for device emulation and guest VM for both KVM and GearV is marked as DVM and GVM respectively. To concretely evaluate the impact of Gear2 on the performance, we build a tailored GearV called G1 which puts the scheduling policy into DVM and removes Gear2. The other well-known embedded hypervisors, such as ACRN, OKL4 and Xvisor, are only used for comparing complexity.

### 4.2. Managed Complexity

Lines of code (LoC) calculated with *cloc* is used as the metrics to evaluate the complexity of hypervisor. Table 4 and Table 5 list the LoC comparison results of hypervisor and device model respectively. GearV is around 10% of the other hypervisors and comparable to ACRN because they belong to the partitioning hypervisor and offload the complex device emulation to user space device model. Thanks to the two gear architecture of GearV, only the partitioning mechanism is required to reside in Gear1, thus its code size can be reduced to two third of ACRN, to further minimize the attack surface of hypervisor and the project scope for certification.

Table 4 Hypervisor LoC Comparison

|  | Xen | KVM | Xvisor | OKL4 | ACRN | GearV |
|---|---|---|---|---|---|---|
| LoC | 309K | 17M | 356K | 393K | 30K | 36K |
| GearV Breakdown |  |  | Gear1 (18K) |  | Gear2 (18K) |  |

Table 5 Device Model LoC Comparison

|  | QEMU | ACRN DM | GDM |
|---|---|---|---|
| LoC | 1M | 43K | 19K |

For the LoC of device model, GearV device model (GDM) is comparable to ACRN DM and much smaller than QEMU used by KVM and Xen, because GDM only implements the virtio based device emulation like serial, network and disk. Other devices like sound card, codec processor, camera sensor and GPU are offloaded to separate user space routines through API forwarding.

### 4.3. Micro-benchmarks

To evaluate the runtime overhead of GearV, we develop five micro-benchmarks within hypervisor to measure the extra time introduced by Gear1 and Gear2, as shown in Table 6. The micro-benchmarks mainly record vmexit information such as reason, frequency and processing duration, so that they can be useful for overhead analysis.

Table 6 Micro-benchmarks

| Case Name | Description |
|---|---|
| Hypercall | HVC-triggered transition between VM and Gear1 |
| World Switch | VCPU transition from one VM to another one. |
| VM trap | Implicit trap from VM to Gear1 hypervisor for device emulation, such as MMIO. |
| IPI | Latency from sending IPI to that the receiving VCPU handles it. |
| I/O out | Latency for IO operation from VM driver, from VM trigger IO operation, to device model handling the request, and acknowledge back |

Table 7 Micro-benchmark Result Comparison

| Micro-benchmark | GearV Guest VM | KVM Guest VM |
|---|---|---|
| Hypercall(ns) | 441 | 3458 |
| VM TRAP(ns) | 732 | 4366 |
| World Switch (ns) | 1485 | 1729* |
| Virtual IPI (ns) | 9928 | 9040 |
| I/O out | 8774 | 7811 |

*world switch performance is not directly tested. We estimate the result by half of hypercall, due to our understanding of KVM that empty hypcall is just two times of world switch.

Table 7 shows the micro-benchmark result comparison between GearV and KVM. The data shows that GearV achieves much better performance than KVM does on hypercall, because hypercall only involves non-volatile register save/restore in Gear1, while it requires world switch from guest VM to its host VM for KVM. The case is similar to VM trap caused by MMIO. GearV achieves similar performance with KVM on world switch, because it requires to save/restore all the register context. For IPI, both GearV and KVM achieve similar performance, because IPI involves a same set of operations on both GearV and KVM, including world switch, interrupt delivery and GIC emulation. The I/O operation has a longer execution path on GearV than it on KVM because it requires more world switch between Gear2 and DVM. Thus, the I/O out takes more time on GearV than KVM.

To further understand the impact of Gear2 on the performance, we analyze the above micro-benchmarks result to estimate the extra overhead introduced by Gear2. As described in section 3.2, the major difference that GearV differs from the traditional monolithic hypervisor is that it introduces Gear2 to handle vcpu scheduling policy triggered by interrupt. So each interrupt handling adds one or two more rounds of world switch. Therefore, if we can collect the interrupt count per second (interrupt frequency), we can estimate the extra overhead introduced by Gear2 through multiplying interrupt frequency by the time of world switch. The interrupt frequency is collected by running CPU and I/O bound benchmark case, as shown in Table 8. And the cost of world switch to Gear2 can be estimated with the formula 1.

Table 8 Interrupt frequency

| Benchmark Case | FIO | Sysbench CPU |
|---|---|---|
| Interrupt count per second | 12346 | 250 |

$$cost = Int_{Freq} * WorldSwith_{cost} * 2 \quad (1)$$

As shown in Table 7, the cost of world switch is 1485ns. Let's round it up to $1.5 * 10^{-6}$ second. Then we can calculate the Gear2-introduced overhead for IO-bound and CPU-bound workload according to Table 8 using the above formula 1.

$$cost\_est_{io} = 12346 * 1.5 * 10^{-6} * 2 = 3.7\%$$
$$cost\_est_{cpu} = 250 * 1.5 * 10^{-6} * 2 = 0.1\%$$

From above estimation, the Gear2-introduced overhead on IO-bound benchmark is around 3.7%, and the one on CPU-bound benchmark is around 0.1%. The total overhead would be slightly higher because there are more overheads besides world switch during interrupt delivery.

### 4.4. Application Benchmarks

*Sysbench*, *FIO* and *GLMark2* are chosen as application benchmarks to evaluate the user-visible performance impact of virtualization techniques on CPU, RAM, I/O and GPU respectively. The same OS and application is run on bare mental, KVM and GearV. For performance comparisons, the native data is treated as the baseline and the VM data is normalized to native data. The detail information and running command of the application benchmarks are shown in Table 9.

Table 9 Application Benchmarks

| Benchmark | Running Command |
|---|---|
| Sysbench 1.1.0 | *CPU: sysbench cpu --threads=2 --cpu-max-prime=200000 run, Memory: sysbench memory run* |
| fio-3.23 | *fio --randrepeat=1 --direct=1 --gtod_reduce=1 --name=test --bs=4k --iodepth=64 --size=100M --readwrite=randrw --rwmixread=75* |
| GLMark2 | *glmark2-es2-wayland --off-screen --size 1280x800* |

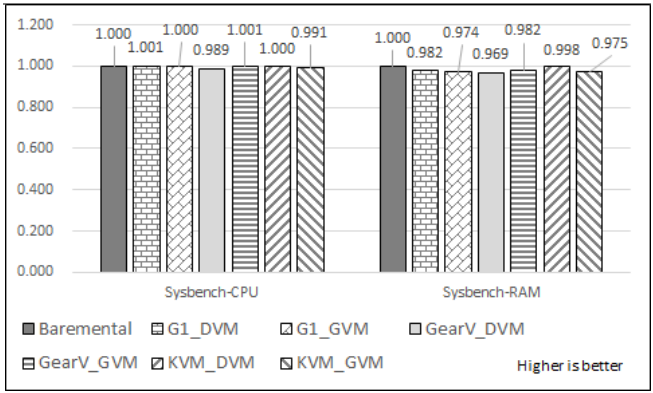

Figure 6 Sysbench Performance Comparison

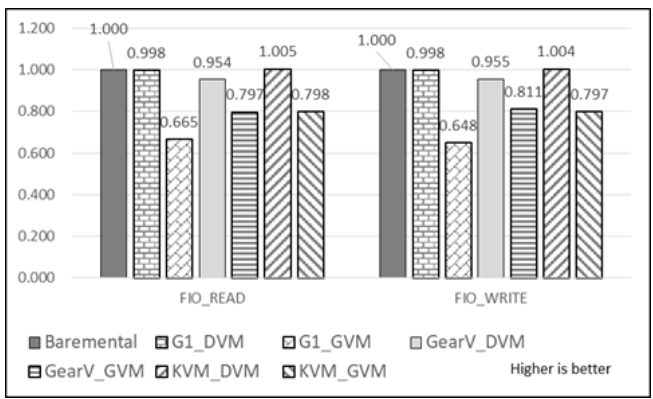

Figure 7 I/O Benchmark

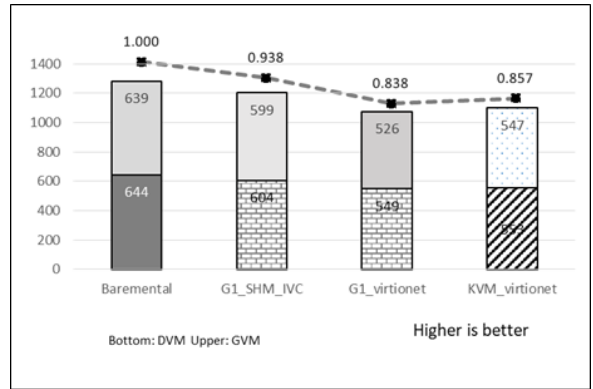

Figure 8 GPU Benchmark

Figure 6 shows the CPU and memory performance comparisons. The result data shows that both CPU and memory performance of VMs is comparable between KVM and GearV, and the virtualization overhead comparing to bare mental is negligible on CPU. The result also shows that the Gear2-introduced overhead is negligible on CPU and memory performance, thanks to that the hardware-assisted virtualization mechanism for CPU and memory is fully utilized in Gear1 and thus it is not necessary to involve Gear2 too much.

Figure 7 shows the I/O performance comparisons. The result shows that the read and write performance of DVM and GVM is also comparable between KVM and GearV. As expected, both G1_DVM and KVM_DVM achieve 100% performance compared with bare metal as they almost run natively. Gear2 introduces 4.5% overhead for DVM (by comparing GearV_DVM and G1_DVM), this is similar to our estimated 3.7% in section 4.3. For secondary VM (GVM), GearV has similar performance with KVM. And we notice that GearV (GearV_GVM) has much better I/O performance than Gear1 (G1_GVM). This is due to low scheduling overhead because Gear2 just binds VCPU to physical CPU.

Figure 8 shows the GPU performance comparisons measured by *GLMark2*. The result data is collected through running one *GLMark2* in DVM and one *GLMark2* in GVM simultaneously, on both KVM and GearV. As mentioned in section 3.7, GPU virtualization is implemented through API forwarding, which intercepts OpenGLES API in GVM and forwards the API commands to DVM via inter-vm communication channel (IVC). The API forwarding based solution is agnostic to hypervisor, thus the GPU virtualization solution can also be ported to KVM. The result shows that the GPU performance is comparable between KVM and GearV, and both achieve around 80% performance comparing to bare mental. The 20% virtualization overhead is mainly caused by the inter-vm communication channel introduced by *virtio-net* which involves data copy based on *virtio*. If change IVC from *virtio-net* to shared memory based IVC, the GPU performance can increase to around 90% of bare mental.

### 4.5. Real Time

To evaluate the real-time performance of GearV hypervisor, the scheduling latency reported by cyclictest is used as the metrics. To collect real-time data, two Linux 4.14 images patched with PREEMPT_RT and Xenomai respectively are treated as RTOS to run above five kinds of platform configurations, including bare mental native, KVM, GearV non-RT VM, GearV RT VM with emulated vGIC and GearV RT VM with pass-through vGIC. Each configuration is run with cyclictest for one hour. Table 10 and Table 11 lists the scheduling latency from cyclictest in microseconds for Xenomai and PREEMPT_RT. The data collected from native run is treated as the baseline jitter. The "Max" column indicates the worst-case latency and we use it to represent the jitter introduced by hypervisor. Therefore, the rightmost column is calculated by normalizing the worst-case latency of RTOS to the one of native run. From the results, the GearV RTVM achieves the best real-time performance, followed by GearV RT VM vGIC. The result data shows that the normalized jitter introduced by GearV is around 5% for Xenomai and 12% for PREEMPT_RT respectively. It proves that the jitter introduced by GearV is negligible compared to native run and static partitioning works best for RTVM because it reduces the vmexit dramatically.

Table 10 Sched latency in microseconds (Xenomai)

| Config | Min | Ave | Max | Normalized Jitter |
|---|---|---|---|---|
| Native | 1 | 2 | 38 | 1 |
| KVM RT VM | 8 | 49 | 3177 | 83.6 |
| GearV non-RT VM | 12 | 67 | 24250 | 638.15 |
| GearV RT VM vGIC | 1 | 11 | 60 | 1.39 |
| GearV RT VM | 1 | 2 | 40 | 1.05 |

Table 11 Sched latency in microseconds (PREEMPT-RT)

| Config | Min | Ave | Max | Normalized Jitter |
|---|---|---|---|---|
| Native | 1 | 6 | 40 | 1 |
| KVM RT VM | 8 | 51 | 4337 | 108.42 |
| GearV non-RT VM | 11 | 66 | 28120 | 703 |
| GearV RT VM vGIC | 1 | 13 | 75 | 1.85 |
| GearV RT VM | 1 | 6 | 45 | 1.12 |

To evaluate the impact of vmexit on real-time performance of GearV, the virtual GIC (vGIC) is implemented in both full emulation mode and pass-through mode, as marked as GearV RT VM vGIC and GearV RT VM respectively in above Table 10 and Table 11. The data shows that the pass-through mode achieves better real-time performance than the full emulation mode, because the frequent vmexit due to GIC emulation are eliminated. Regarding to the 5%~12% jitter overhead of GearV RT VM, we suspect that they are caused by the contention on LLC and memory bus bandwidth, because a busy loop writing to disk is running in the background when run *cyclictest*. The cache coloring and memory bus bandwidth reservation is developed in progress during this paper writing, thus we leave its evaluation as the future work.

### 5. Conclusions and Future Work

With the rapid adoption of powerful computing platform in IoT such as industrial control and automotive scenarios, there is a strong demand to better utilize hardware resources while fulfilling the stringent real-time and reliability requirements regulated by certification. Embedded virtualization is an appealing technology to fulfill the above goals through consolidating systems with different criticality level on a single computing platform. The general approach to achieve stringent reliability is to use the static partitioning hypervisor, but it reveals limitation on resource efficiency. The widely-used server grade hypervisors such as KVM and Xen do well on resource efficiency, but they were not designed with embedded constraints in mind and restructuring them to fulfill the stringent reliability requirement involves huge engineering efforts. We argue that stringent reliability and resource efficiency don't always match, and favoring one may break the other. We propose GearV, a two-gear hypervisor architecture to divide the functionality of hypervisor into partitioning mechanism and scheduling policy, which is implemented in Gear1 and Gear2 respectively. In this way, Gear1 is a self-contained type-1 hypervisor and it is easier for Gear1 to get certification on system reliability. Gear2 runs in the primary VM of Gear1 to handle complex logics related to resource efficiency, such as vcpu scheduling and interrupt processing. We extend Hafnium to implement a reference of GearV and evaluate it according to the design goals on managed complexity, virtualization overhead and real-time performance. The result data shows that LoC of Gear1 is around 18K, two third of the smallest one in the open-source community. The extra overhead introduced by Gear2 is negligible on CPU and memory virtualization, and 4.5% on I/O virtualization. GearV supports real-time VM with only 5% jitter overhead measured by Xenomai. We believe that GearV can serve as a reasonable hypervisor architecture for the mixed-criticality IoT systems. The evaluation demonstrates that GearV could fulfill the design goals on managed complexity, reliability and real-time with acceptable overhead, and the two-gear architecture can be easily applied to other open-source hypervisors.

As for the future work, there are several areas. First, the microarchitecture level partitioning techniques, such as cache coloring and memory bus bandwidth reservation, can be explored to enhance the real-time performance of GearV. Second, new vcpu scheduling policies can be implemented in Gear2 to explore the optimization opportunities on resource efficiency according to the workloads adaptively. Last, with the guidance of two-gear architecture, other open source hy-pervisors can be restructured with acceptable engineering effort to fulfill the stringent reliability and resource efficiency requirement of mixed-criticality IoT systems.